# Waveform Design Using Half-duplex Devices for 6G Joint Communications and Sensing

Yihua Ma[1,2], Zhifeng Yuan[1,2], Guanghui Yu[1,2], Shuqiang Xia[1,2], Liujun Hu[1,2]
[1]ZTE Corporation, Shenzhen, China
[2]State Key Laboratory of Mobile Network and Mobile Multimedia Technology, Shenzhen, China
Email: {yihua.ma, yuan.zhifeng, yu.guanghui, xia.shuqiang, hu.liujun}@zte.com.cn

*Abstract*—Joint communications and sensing is a promising 6G technology, and the challenge is how to integrate them efficiently. Existing frequency-division and time-division coexistence can hardly bring a gain of integration. Directly using orthogonal frequency-division multiplexing (OFDM) to sense requires complex in-band full-duplex to cancel the self-interference (SI). To solve these problems, this paper proposes novel coexistence schemes to gain super sensing range (SSR) and simple SI cancellation. SSR enables JCS to gain a sensing range of a sensing-only scheme and shares the resources with communications. Random time-division is proposed to gain a super Doppler range. Flexible sensing implanted OFDM (FSI-OFDM) is also proposed. FSI-OFDM uses random sensing occasions to gain super Doppler range, as well as utilizes the fixed tail sensing occasions to achieve supper distance range. The simulation results show that the proposed schemes can gain SSR with limited resources.

## I. INTRODUCTION

6G is going to not only evolve in terms of spectral efficiency, latency, and connectivity but also seek to provide beyond-communications services. Joint communications and sensing (JCS) [1]-[3] is proposed to provide sensing services through the communications devices. The RF convergence of these two functions also makes it possible to realize an efficient joint scheme to share the resources including spectrum and hardware.

Although the unified design is always preferable to save the cost, the two functions themselves have different working principles. Communications aim to get the information from the transmitted signal itself while sensing focuses on the channel information. Communications usually employ orthogonal frequency-division multiplexing (OFDM), as it provides robustness against multi-path channels, simple equalization and flexible resource allocation. In radar sensing, the widely used solution is based on frequency modulated continuous wave (FMCW) or chirp signal for its large bandwidth, simple processing scheme, and importantly, simple self-interference (SI) cancellation.

Some papers [2] [4] suggested using OFDM to sense. The data transmission efficiency and flexibility can be ensured, and the sensing overheads can be reduced via reusing data symbols to sense. The problem is that a complex in-band full-duplex transceiver is required. As SI is much stronger than the echos, Ful-duplex usually cancels SI in multiple domains, including spatial domain, RF/analog domain and digital domain. When the multiple-input and multiple-output (MIMO) system is used, all transmit antennas generate SI [5], which makes the SI cancellation much more complex than the single-antenna situation.

FMCW was also considered to communicate in JCAS. The simplest way is to modulate the amplitude, frequency or phase of the chirp signal, which is only for low-rate communications. OFDM chirp methods were designed to generate orthogonal FMCW signals for MIMO radar [7]. Furthermore, Orthogonal chirp division multiplexing (OCDM) replaces the Fourier transform kernel in OFDM with the Fresnel transform [8] and uses a DFT-spread-OFDM (DFT-s-OFDM) receiver. Although FMCW and OFDM are combined, these methods lose the advantages of multi-path robustness of OFDM, as well as the efficient SI suppression of FMCW.

To achieve the advantages of both OFDM and FMCW. one way is to keep them orthogonal. However, if JSAC uses time or frequency division, there is no spectrum efficiency gain compared to separated systems. This paper proposes to use random sampling with limited sensing resources to gain a sensing range as large as that all resources are used for sensing, which is named super sensing range (SSR). Random time-division (RTD) is proposed to gain a super range of Doppler. To make it more compatible with existing OFDM systems, flexible sensing implanted OFDM (FSI-OFDM) is proposed. FSI-OFDM maps the time-domain chirp signals to different sub-carriers and is equivalent to a short spreading of sub-carriers. A special spreading code set is designed to ensure time-domain randomness. The spreading code can be either randomly selected to gain a super Doppler range or fixed at the symbol tail to gain a super distance range with the help of cyclic prefix (CP). Moreover, the SI of both sensing and data signals are canceled in simple analog hardware. Finally, simulations are provided to verify the proposed SSR schemes.

The contributions of this paper include: (1) RTD and FSI-OFDM are proposed to gain SSR and save the resources, (2) the mapping from generalized time-domain repeated signals to the sub-carriers of OFDM is provided and proved; (3) hardware-friendly algorithm for FSI-OFDM is proposed with simple analog processing to cancel SI without the full-duplex requirement. In the rest of this paper, Section II briefly summarizes existing works, Section III introduces the proposed RTD and FSI-OFDM, and Section IV gives some simulation results. In this paper, $(\cdot)^*$, $(\cdot)^T$, and $(\cdot)^H$ denote conjugate, transpose, and Hermitian transpose of a matrix or vector, respectively. $\otimes$, $\circ$ and mod represent the Kronecker product, Hadamard product and modulo operation.

## II. CONVENTIONAL METHODS

### A. OFDM-based Methods

OFDM is good for communications as it supports a flexible resource allocation and converts the convolution channel into a dot-product channel. It can be expressed as

$$\mathbf{x} = \mathbf{F}_N^{-1}\mathbf{s}, \quad (1)$$

where $\mathbf{x} \in \mathbb{C}^{N\times 1}$, $\mathbf{F}_N \in \mathbb{C}^{N\times N}$, and $\mathbf{s} \in \mathbb{C}^{N\times 1}$ denote the time-domain data vector, the normalized $N$-dimensional discrete Fourier transform (DFT) matrix, and the frequency-domain data vector. $N$ is the number of inverse fast Fourier transform points in OFDM. The information bits are modulated and encoded in $\mathbf{s}$, while $\mathbf{x}$ is transmitted in the air.

As mentioned before, some works [2] [4] proposed to directly use OFDM as the JCAS waveform. The main challenge is that OFDM sensing requires a full-duplex transceiver. Full-duplex provides a 2-fold throughput, which is very attractive. However, it is still very challenging to implement full-duplex in practice with overall consideration of complexity, device size and cost. Although some works proposed to use the spatial degree of freedom to realize a full-duplex with a low cost [6], it cannot be used for OFDM sensing as the simultaneous uplink and downlink are in different beams. MIMO makes full-duplex more challenging to realize as the SI is from all transmit antennas [5]. The common half-duplex communications scenario is assumed, and thus OFDM sensing is not considered.

*B. FMCW-based Methods*

FMCW is widely used in radar, and it usually uses a large bandwidth to increase the time-domain resolution. The advantages include low cost, low PAPR, and strong SI suppression ability. One typical form of FMCW signal is the chirp signal as

$$x_n = \exp\left(j2\pi\left(f_0 n T_S + \frac{1}{2} k_c n^2 T_S^2\right)\right), n = 0,1,...,N_C\text{-}1, \quad (2)$$

where $f_0$ is the chirp start frequency, $T_S$ is the sampling time interval, $k_c$ is the chirp rate, and $N_S$ is the number of sampling points. The time width of a chirp is $T_{chirp} = N_C T_S$. Although the echo signal is much weaker than SI, they can be separated in the frequency domain after a mixer. FMCW receiver converts the target signal with delay to the intermediate frequency (IF), while SI is into zero frequency and can be easily filtered. The frequency of the IF signal is the echo delay multiplied by a factor of $k_c$. Since the SI cancellation is done in the analog domain, the dynamic range of echo signals after the quantization can be ensured.

To increase the throughput of chirp transmission, an OFDM-like structure [8] was also proposed, which replaces the Fourier transform kernel with the Fresnel transform kernel. A receiver scheme similar to DFT-s-OFDM can be used. [9] further extended this work to a DFT-s-OFDM transceiver, and employs circularly-shifted orthogonal chirps. However, these methods use a dimension of DFT in spreading as large as that of inverse DFT in OFDM, while DFT-s-OFDM is usually used for a narrowband signal. It becomes neither robust against multi-path channels nor flexible to allocate the resources. Also, the simple SI suppression method can no longer be used as there are a bunch of circularly-shifted chirps, and the mixer cannot convert all SI of them into the zero frequency.

### III. THE PROPOSED SSR SCHEMES

The overall structure of the proposed SSR schemes is in Fig. 1. The sensing function is deployed at the base station (BS). At BS, the transmitter is shared, while the receivers are dedicated. Compared to widely deployed half-duplex BS, the only hardware modification is to add an FMCW receiver.

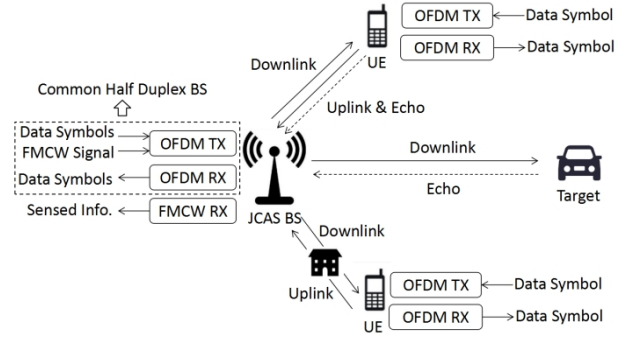

Fig. 1. The overall structure of the proposed SSR JCAS schemes

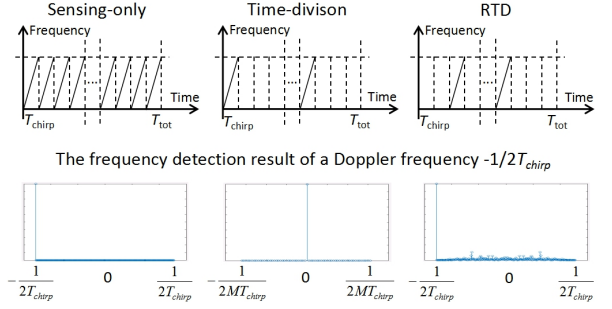

Fig. 2. The Doppler sensing performance comparison of sensing-only, time-division and RTD schemes.

*A. Basic Idea and RTD*

First, time-division coexistence is considered. Compared to frequency-division coexistence, time-division can use all the bandwidth for sensing. Assume that the total time is $T_{tot} = MKT_{chirp}$, where $M$ is the number of chirps/slots in one time group, and $K$ is the number of time groups. The chirp has a length of one time slot. If all resources are used for sensing, the sensed Doppler frequency range is $[-1/2T_{chirp}, 1/2T_{chirp})$. If the first slot in every time slot group is used for sensing, the Doppler range is reduced to $1/M$ due to the sampling frequency reduction. As shown in Fig. 2, the frequency out of the range is aliased into the lower frequency. That is to say, the Doppler range is reduced with partial time-division resources in the existing method.

Compared to existing periodic time-division coexistence, RTD randomly samples the Doppler frequency. In Fig. 2, $K$ slots out of $MK$ slots are randomly selected for sensing, and the sensing resources used are still $1/M$ of the total resources. RTD is still able to recover the Doppler frequency ranging as large as the sensing-only schemes via simple matched filtering. Compared to sensing-only schemes, the left $(M-1)/M$ resources can be used for communications. It can be seen from Fig. 2 that the power of target Doppler frequency is much stronger than the cross-frequency interference, which means that RTD is effective to gain a super Doppler range with only partial resources.

*B. FSI-OFDM Waveform*

The data transmission occasions in RTD are neither continuous nor periodic, which is not friendly to scheduling and low latency in communications. This paper further proposes to implant the chirp signals into the OFDM symbol, which ensures that the data transmission is still regular. A simple implanting case is first introduced, where the chirp signal is repeated for $M$ times in the time domain, and in this case, it only uses $1/M$ of the total $N = MN_{Chirp}$ sub-carriers in one OFDM symbol. Single chirp has $N/M$ sampling points and is written as $\mathbf{p}_C = [p_1, p_2, ..., p_{N/M}] \in \mathbb{C}^{N/M \times 1}$. $M$ repeated chirps in one OFDM symbol form a large vector of $\mathbf{x}_C = \mathbf{1}_{M \times 1} \otimes \mathbf{p}_C \in \mathbb{C}^{N \times 1}$. The corresponding frequency-domain signal $\mathbf{s}_F$ is

$$\mathbf{s}_C = \mathbf{F}_N \mathbf{x}_C = \mathbf{F}_N(\mathbf{1}_{M \times 1} \otimes \mathbf{p}_C) = \mathbf{F}_N(\mathbf{1}_{M \times 1} \otimes \mathbf{I}_{N/M})\mathbf{p}_C. \quad (3)$$

Define that $\mathbf{A} = \mathbf{F}_N(\mathbf{1}_{M \times 1} \otimes \mathbf{I}_{N/M}) \in \mathbb{C}^{M \times N}$. Its element in the $i$-th row and $j$-th column is

$$a_{i,j} = \frac{1}{\sqrt{N}} \sum_{m=1,...,M} e^{j\frac{2\pi}{M}(i-1)(m-1)} e^{j\frac{2\pi}{N}(i-1)(j-1)},$$
$$i = 1,...,N, j = 1,...,M. \quad (4)$$

When $(i-1) \bmod M = 0$,

$$a_{i,j} = \frac{M}{\sqrt{N}} e^{j\frac{2\pi}{N}(i-1)(j-1)} = \sqrt{M} \frac{1}{\sqrt{N/M}} e^{j\frac{2\pi}{N/M}(h-1)(j-1)},$$
$$h = \frac{i-1}{N} + 1 = 1,...,M, j = 1,...,M. \quad (5)$$

Equation (5) shows that $\mathbf{A}$ is a normalized $N/M$-dimensional Fourier transform with a factor of $\sqrt{M}$. When $(i-1) \bmod M \neq 0$,

$$a_{i,j} = \frac{1 - e^{j 2\pi(i-1)}}{1 - e^{j\frac{2\pi}{M}(i-1)}} e^{j\frac{2\pi}{N}(i-1)(j-1)} = 0 \cdot e^{j\frac{2\pi}{N}(i-1)(j-1)} = 0. \quad (6)$$

Therefore,

$$\mathbf{s}_C = \text{vec}\left(\begin{bmatrix} \sqrt{M}(\mathbf{F}_{N/M}\mathbf{p}_C)^T \\ \mathbf{0}_{(M-1) \times N/M} \end{bmatrix}\right). \quad (7)$$

where vec(·) means the vectorization operation. That is to say, $M$ repeated FMCW signal only takes up the first sub-carrier every $M$ sub-carriers, while the left $(M-1)/M$ can be used for data transmission. A generalized case is for

$$\mathbf{x}_{Cm} = (\mathbf{1}_{M \times 1} \otimes \mathbf{p}_C) \circ \mathbf{r}_m, m = 1,...,M,$$
$$\mathbf{r}_m = [r_{m1} \ r_{m2} \ ... \ r_{mN}]^T, r_{mn} = e^{j\frac{2\pi}{N}(m-1)(n-1)}. \quad (8)$$

Using similar steps from (3) to (7), it is easy to get

$$\mathbf{s}_{Cm} = \text{vec}\left(\begin{bmatrix} \mathbf{0}_{(m-1) \times N/M} \\ \sqrt{M}(\mathbf{F}_{N/M}\mathbf{p}_C)^T \\ \mathbf{0}_{(M-m) \times N/M} \end{bmatrix}\right). \quad (9)$$

Obviously, $\mathbf{x}_{Cm}$ for different $m$ has a constant envelope in the time domain. To increase the randomness of Doppler sampling, a DFT-like spreading code set is designed as

$$\mathbf{B} = \begin{bmatrix} \mathbf{b}_1^T \\ \mathbf{b}_2^T \\ ... \\ \mathbf{b}_M^T \end{bmatrix} = \left(F_M \circ \left(\mathbf{1}_{M \times 1} \otimes \begin{bmatrix} e^{-j\frac{\pi}{M} \cdot 0} \\ e^{-j\frac{\pi}{M} \cdot 1} \\ ... \\ e^{-j\frac{\pi}{M}(M-1)} \end{bmatrix}^T\right)\right) \begin{bmatrix} \mathbf{x}_{c1}^T \\ \mathbf{x}_{c2}^T \\ ... \\ \mathbf{x}_{cM}^T \end{bmatrix}$$
$$= \mathbf{U}\mathbf{X}_{\text{Base}}, \mathbf{U} = [\mathbf{u}_1, \mathbf{u}_2, ..., \mathbf{u}_M]^T. \quad (10)$$

where $\mathbf{u}_m = [u_{m,1}, u_{m,2}, ..., u_{m,M}]^T \in \mathbb{C}^{M \times 1}$ is the combination coefficient vector to combine rows in $\mathbf{X}_{\text{base}}$. As equation (9) shows, each row vector in $\mathbf{X}_{\text{Base}}$ has a frequency-domain counterpart with the same values at each corresponding sub-carrier. Therefore, $\mathbf{u}_m$ can also be seen as the sub-carrier-wise spreading code. In FSI-OFDM, $\mathbf{b}_m$ is randomly selected in each OFDM symbol. There are ($M$-1) orthogonal spreading codes left for data transmission. The transmit signal of one OFDM symbol is

$$\mathbf{x} = \mathbf{b}_m + \mathbf{F}_N^{-1} \sum_{i \neq m}(\mathbf{d}_i \circ \mathbf{u}_i), \quad (11)$$

where $\mathbf{d}_i \in \mathbb{C}^{N/M \times 1}$ is the $i$-th frequency domain transmit data signal.

The chirp signal in the $\mathbf{b}_m$ is compared with that in RTD in Fig. 3. The frequency of chirp signal in one time occasion is from -$B$/2 to $B$/2, where $B$ is the bandwidth of one OFDM symbol. As $M$ is usually very small in FSI-OFDM, the spreading has little impact on the channel features of OFDM symbols. There can also be multiple $\mathbf{b}_m$ being selected in one OFDM symbol, but this paper only discusses the one $\mathbf{b}_m$ situation. FSI-OFDM implants chirp in the OFDM and is thus compatible with the current communications schemes.

*C. Fully Analog SI Cancellation*

With FSI-OFDM, FMCW and data are now orthogonal in the code domain. In the OFDM receiver, it is easy to separate these two parts via despreading M continuous sub-carriers. However, in the FMCW receiver, such digital operations cannot be used, as the dynamic range of echo signal will be greatly affected if SI is not canceled before quantization. This paper proposes fully analog SI cancellation in Fig. 4.

The mixer and filter are used to cancel the SI of the chirp signal, which is widely used in the FMCW radar system. Moreover, the mixer automatically provides a multiplication of conjugate time-domain coefficients for chirps. The SI signal is the same as the transmit signal in (11), which means $\mathbf{x}_{SI} = \mathbf{x}$. The mixer operation for the SI signal gets

$$\mathbf{y}_{\text{SIbeat}} = \mathbf{b}_m \circ \left(\mathbf{b}_m + \sum_{i \neq m} \mathbf{F}_N^{-1}(\mathbf{d}_i \circ \mathbf{u}_i)\right)^*$$
$$= \mathbf{b}_m \circ \mathbf{b}_m^* + \mathbf{b}_m \circ \left(\sum_{i \neq m} \mathbf{F}_N^{-1}(\mathbf{d}_i \circ \mathbf{u}_i)\right)^* = \mathbf{y}_{\text{SIbeat1}} + \mathbf{y}_{\text{SIbeat2}}, \quad (12)$$

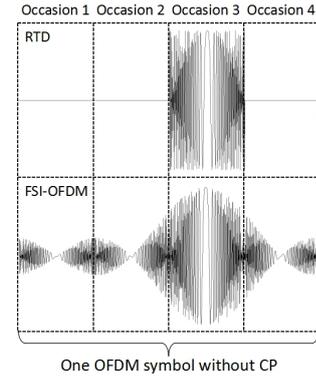

Fig. 3. The real part of the chirp signal in RTD and FSI-OFDM ($M$ = 4).

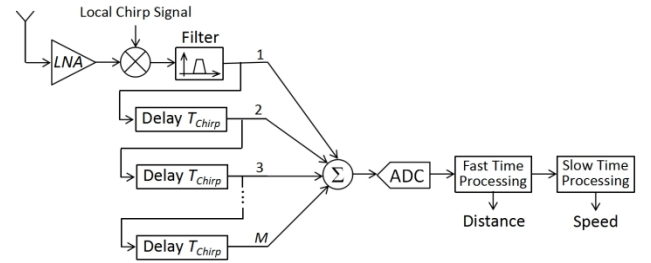

Fig. 4. The block diagram of the FMCW receiver for FSI-OFDM.

The first item $\mathbf{y}_{SIbeat1}$ is at zeros frequency and can be easily filtered, while the second item $\mathbf{y}_{SIbeat2}$ is removed using other operations. As proved in the previous text, the time domain $\mathbf{b}_m$ is

$$\mathbf{b}_m = \mathbf{u}_m^T \mathbf{X}_{Base} = \sum_{i=1}^{M} u_{m,i}(\mathbf{1}_{M\times 1}\otimes \mathbf{p}_C)\circ \mathbf{r}_i = \sum_{i=1}^{M} u_{m,i}\mathbf{r}_i'\otimes(\mathbf{p}_C\circ\mathbf{r}_i''),$$

$$\mathbf{r}_i' = \mathbf{r}_i\left[1, \frac{N}{M}+1,\ldots,(M-1)\frac{N}{M}+1\right], \mathbf{r}_i'' = \mathbf{r}_i\left[1,2,\ldots,\frac{N}{M}\right]. \quad (13)$$

The frequency-domain counterpart of $\mathbf{b}_m$ is

$$\mathcal{F}(\mathbf{b}_m) = \mathbf{F}_N \mathbf{b}_m = \sqrt{M}\mathbf{F}_{M/N}\mathbf{p}_C\circ \mathbf{u}_m. \quad (14)$$

As in the derivation, the property of chirp is not used, and the relationship between equation (13) and (14) also work for data signals. $\mathbf{y}_{SIbeat2}$ can be written as

$$\mathbf{y}_{SIbeat2} = \left(\sum_{k=1}^{M} u_{m,k}\mathbf{r}_k'\otimes(\mathbf{p}_C\circ\mathbf{r}_k'')\right)\circ\left(\sum_{i\neq m}\sum_{j=1}^{M} u_{i,j}\mathbf{r}_j'\otimes\left(\mathbf{F}_{N/M}^{-1}\mathbf{d}_i\circ\mathbf{r}_j''\right)\right)^*$$

$$= \sum_{i\neq m}\sum_{j=1}^{M}\sum_{k=1}^{M}\left((u_{m,k}\mathbf{r}_k')\circ(u_{i,j}\mathbf{r}_j')^*\right)\otimes\left((\mathbf{p}_C\circ\mathbf{r}_k'')\circ\left(\mathbf{F}_{N/M}^{-1}\mathbf{d}_i\circ\mathbf{r}_j''\right)^*\right)$$

$$= \sum_{i\neq m}\sum_{j=1}^{M}\sum_{k=1}^{M}\left(u_{m,k}u_{i,j}^*\mathbf{r}_k'\circ\mathbf{r}_j'^*\right)\otimes\left(\mathbf{p}_C\circ\left(\mathbf{F}_{N/M}^{-1}\mathbf{d}_i\right)^*\circ(\mathbf{r}_k''\circ\mathbf{r}_j''^*)\right)$$

$$= \sum_{i\neq m}\sum_{t=0}^{M-1}\sum_{k=1}^{M}\left(u_{m,k}u_{i,j}^*\mathbf{r}_t'\right)\otimes\left(\mathbf{p}_C\circ\left(\mathbf{F}_{N/M}^{-1}\mathbf{d}_i\right)^*\circ\mathbf{r}_t''\right),$$

$$j = (k-t)\bmod M. \quad (15)$$

After the delay and sum operation, equation (15) becomes

$$\mathbf{y}_{SIsum} = \sum_{i\neq m}\sum_{t=0}^{M-1}\sum_{k=1}^{M}\sum_{l=1}^{M}\left(u_{m,k}u_{i,j}^*\mathbf{r}_t'[l]\right)\left(\mathbf{p}_C\circ\left(\mathbf{F}_{N/M}^{-1}\mathbf{d}_i\right)^*\circ\mathbf{r}_t''\right)$$

$$= \sum_{i\neq m}\sum_{t=0}^{M-1}\left(\mathbf{p}_C\circ\left(\mathbf{F}_{N/M}^{-1}\mathbf{d}_i\right)^*\circ\mathbf{r}_t''\right)\sum_{k=1}^{M}\sum_{l=1}^{M}u_{m,k}u_{i,j}^*\mathbf{r}_t'[l],$$

$$j = (k-t)\bmod M. \quad (16)$$

For $t = 0, j = k$. As $\mathbf{u}_m$ and $\mathbf{u}_i$ are orthogonal,

$$\sum_{k=1}^{M}\sum_{l=1}^{M}\left(u_{m,k}u_{i,k}^*\mathbf{r}_0'[l]\right) = \sum_{l=1}^{M}(\mathbf{r}_0'[l])\sum_{k=1}^{M}u_{m,k}u_{i,k}^* = \sum_{l=1}^{M}(\mathbf{r}_0'[l])\cdot 0 = 0. \quad (17)$$

For $t\neq 0$, as $\mathbf{r}_t'$ is orthogonal to $\mathbf{1}_{M\times 1}$,

$$\sum_{k=1}^{M}\sum_{l=1}^{M}\left(u_{m,k}u_{i,j}^*\mathbf{r}_t'[l]\right) = \sum_{l=1}^{M}(\mathbf{r}_t'[l])\sum_{k=1}^{M}u_{m,k}u_{i,j}^* = 0\cdot\sum_{k=1}^{M}\left(u_{m,k}u_{i,j}^*\right) = 0. \quad (18)$$

Therefore, $\mathbf{y}_{SIsum} = 0$. The SI of both chirp and data signals are canceled by simple analog operations.

*D. Super Range of Doppler and Distance*

FSI-OFDM uses the same idea as RTD to gain super Doppler range. As shown in Fig. 3, the chirp signal in FSI-OFDM is very similar to that in the RTD. Moreover, to simplify the processing, FSI-OFDM adopts the same matched filtering method used in RTD. That is to say, the chirp signal in FSI-OFDM is viewed as its counterpart in RTD. The only difference is that the position of chirp in FSI-OFDM is not arbitrary. Assume there are $K$ OFDM symbols, and in the $k$-th OFDM symbols, the $\alpha_k$-th code, corresponding to the $\alpha_k$-th slot of every $M$ slots in RTD, is used for sensing. The matched filter matrix is

$$\mathbf{F}_{M\times K}^H = \left[\mathbf{f}_{\alpha_1}, \mathbf{f}_{M+\alpha_2}, \ldots, \mathbf{f}_{(K-1)M+\alpha_K}\right]^H, \quad (19)$$

where $\mathbf{f}_a$ is the $a$-th column vector of $\mathbf{F}_{MK}$.

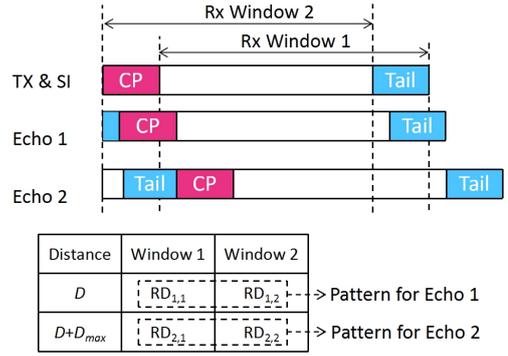

Fig 5. Using the CP to add a new receiving window to distinguish whether the distance is beyond the range.

Till this step, the CP of OFDM has not been mentioned for simplicity. CP is a replica of the tail of the time-domain OFDM symbol and ensures the orthogonality of sub-carriers when the delay of multi-path is within the CP length. The sensing signal does not require such a CP, and the CP cost can be a resource waste. Assume that the point number of CP is $N_{CP}$. The receiving window only receives $N$ points for one OFDM symbol, which means the utilization ratio is $N/(N+N_{CP})$. Fortunately, as sensing focuses on channel information instead of transmitted bits, CP can be utilized to get the sensing information.

Fig. 5 shows how super distance range works. In this case, $\mathbf{u}_M$ is used for sensing to focus the sensing energy at the symbol tail. As CP is the replica of the symbol tail, CP also gains large sensing energy. Two receiving windows can be used. The orthogonality of sub-carriers in SI signal can be ensured, and the SI cancellation processing works for both windows. It does not matter that the sub-carriers are no longer orthogonal in window 2, as the interference of the data echo signal is very weak after slow time processing. The ideal way to utilize the pattern is to recover the delay and Doppler using super-resolution methods and reconstruct the two patterns to see which one is similar.

As window 2 detects the tail of the previous symbol, a rotation is added for every symbol to ensure phase continuity. To be specific, the $k$-th symbol is multiplied by a factor of $e^{j2\pi(k-1)/M}$. Also, a simple method is used to get the super Doppler range. A pattern matrix $\mathbf{P}\in\mathbb{C}^{N/M\times K\times 2\times 2}$ can be constructed with its element $p_{abcd}$ representing the response of $a$-th delay and $b$-th Doppler with echo $c$ in the window $d$. Then, inverse the 2x2 matrix for every $a$ and $b$, the matrix $\mathbf{P}_{sol}\in\mathbb{C}^{N/M\times K\times 2\times 2}$ can be used to solve the pattern. $\mathbf{P}_{sol}$ is used to linearly combine the elements of two $N/M\times K$ range-Doppler (RD) matrices. In simulations, the linear combination vectors in the inverse matrix are normalized.

IV. NUMERICAL RESULTS

*A. Simulation settings*

In simulations, the carrier frequency is 60Ghz and the sub-carriers spacing of OFDM, $B_{SCS}$, is 60kHz. $N = 2048$, and the guard band is not considered, which means the total bandwidth $B$ is 122.88MHz. The sampling time interval of both the chirp signal and data signal is 8.118 ns. Extended CP mode is considered as this mode has a fixed length of CP. That is to say, $N_{CP} = 512$. $M$ is set to be 4. The time of one chirp is 4.167 us, which is fixed among all schemes. For sensing-only, time-division and RTD schemes, $K = 80$. As FSI-OFDM has extra CP, $K$ is set to 64 for the same total time length. That is to say, the Doppler resolution of FSI-OFDM is slightly more coarse than other schemes due to the CP required by OFDM. The echo signal is assumed to be 100 dB lower than the SI without considering the path loss model, and the SNR of the echo signal is -10 dB.

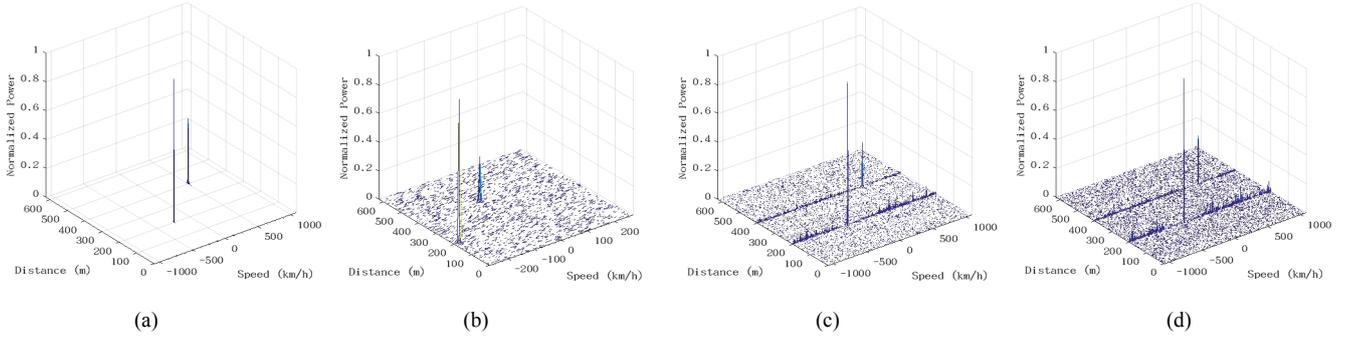

(a)                 (b)                 (c)                 (d)

Fig. 6. The mesh surfaces of the RD matrix for (a) sensing-only, (b) time-division, (c) RTD, and (d) FSI-OFDM schemes. The power is normalized by the maximum value. The elements with normalized power less than $10^{-3}$ are not displayed to make the diagram clear.

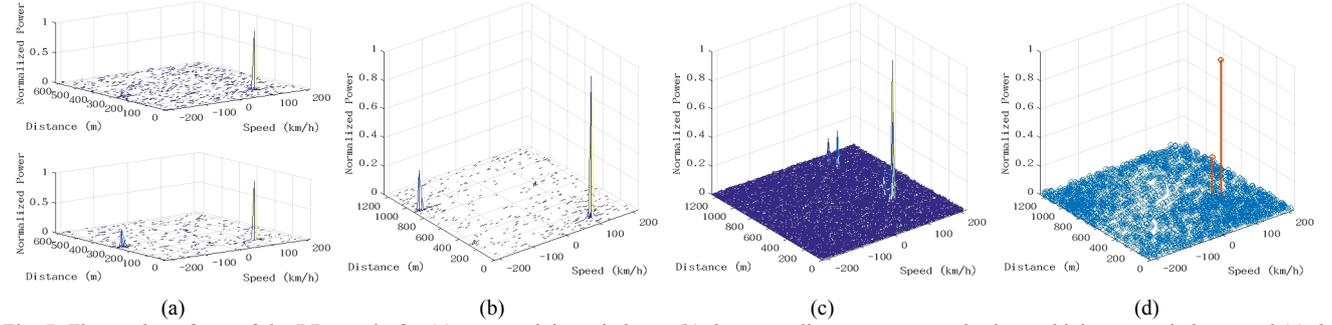

(a)                 (b)                 (c)                 (d)

Fig. 7. The mesh surfaces of the RD matrix for (a) two receiving windows, (b) the super distance range result via combining two windows, and (c) the super distance range result with errors; and (d) the RD discrete points of the correct super distance range result after removing values around the peaks.

*B. Super Doppler Range*

To show super Doppler range, two targets of (200m, -250km/h) and (400m, 500km/h) are assumed. For the sensing-only schemes, as the distance and speed/Doppler are within its range, two targets can be easily detected as shown in Fig. 6(a). For conventional time-division schemes using $1/M$ resources for sensing, a speed of 500km/h is beyond its Doppler range of ±270 km/h. A wrong result of (400m, -40 km/h) is detected in Fig. 6(b). As a comparison, RTD and FSI-OFDM successfully detect two targets in Fig. 6(c) and Fig. 6(d) with the same amount of resources. The interference in FSI-OFDM is stronger than that in RTD, as the useful resource percentage is reduced to $N/(N+N_{CP})/M$ with extra overheads for CP. The super Doppler range is also verified in other cases, and the results are similar to those shown in Fig. 6.

*C. Super Distance Range*

Fig. 7 shows the result of the super distance range. Two targets have profiles of (100 m, 100 km/h) and (900 m, -100 km/h), respectively. In Fig. 7(a), the result in each window is wrong as the object at 900m is out of range. Using the simple linearly combination method, a super distance range result is obtained in Fig. 7(d), which successfully detects two targets. We have to admit that this simple method only works perfectly in the on-grid RD results. As shown in Fig. 7(c), when two targets are (400 m, 100 km/h) and (500 m, 100 km/h), the values around peaks cannot be perfectly combined.

It can be easily solved by only keeping the peak values for each target while forcing the surrounding values to be zeros in each window. Then, Fig. 7(d) can be obtained, which correctly detects the two targets.